\RequirePackage{ifpdf}
\documentclass{PoS}

\usepackage[pdftex]{graphicx}
\usepackage[font={small,it},labelfont=bf,format=hang,labelsep=colon,justification=raggedright,width=.9\textwidth]{caption}
\usepackage{subcaption}

\title{
	Transverse momentum fluctuations and $\mathsf{\Delta\eta\Delta\phi}$ correlations
	in p+p interactions at the CERN SPS energies
}

\ShortTitle{
	$p_{T}$ fluctuations and $\mathsf{\Delta\eta\Delta\phi}$ correlations
	in p+p at the CERN SPS energies
}

\author{
	\speaker{Tobiasz Czopowicz} and Bartosz Maksiak for the  NA61/SHINE Collaboration \\
	Warsaw University of Technology \\
	E-mail: \email{tobiasz.czopowicz@cern.ch}
}

\abstract{
	The NA61/SHINE experiment aims to discover the critical point of strongly interacting
	matter and study the properties of the onset of deconfinement. These goals are to be
	achieved by performing a two dimensional phase diagram ($T-\mu_B$) scan by measurements
	of hadron production properties in proton-proton, proton-nucleus and nucleus-nucleus
	interactions as a function of collision energy and system size. Close to the critical
	point an increase of fluctuations is predicted. \\
	In this contribution preliminary results on transverse momentum fluctuations
        and two-particle pseudorapidity/azimuthal angle correlations
	from the NA61/SHINE energy scan of p+p collisions will be presented. These new
	results will be compared with NA49 data on central Pb+Pb collisions and model predictions.
}

\FullConference{
	Critical Point and Onset of Deconfinement \\
	March 11-15 2013 \\
	Napa, CA, USA
}

\begin{document}

	\section{Introduction}
	The NA61/SHINE experiment aims to discover the critical point of strongly interacting
	matter and study the properties of the onset of deconfinement. Close to the phase
	transition and close to the critical point large fluctuations are predicted.

	This poster presents results from the analysis of transverse momentum event-by-event fluctuation
	analysis and two-particle correlations.

	\section{The $\mathsf{\Phi}$ measure}
	$\Phi$ \cite{Phi_measure} is a strongly intensive measure of fluctuations.
	In the Wounded Nucleon Model it does not depend on the number of wounded nucleons nor on its fluctuations.
	In thermodynamical models $\Phi$ does not depend on the volume and its fluctuations. \\
	$\Phi$ was used extensively by NA49 to study transverse momentum fluctuations ($\Phi_{p_{T}}$)
	in Pb+Pb collisions \cite{NA49}.

	The $\Phi_{p_T}$ formula used for the analysis can be expressed by event quantities:
	\begin{center}
		$\Phi_{p_T} \equiv \sqrt{\frac{\langle X^{2} \rangle}{\langle N \rangle } -
		\frac{2  \langle X \rangle \; \langle NX \rangle}{ {\langle N \rangle}^2 } +
		\frac{ {\langle X \rangle}^2 \;   \langle N^{2} \rangle}{ {\langle N \rangle }^3 } } -
		\sqrt{ \frac{\langle X_{2} \rangle}{{\langle N \rangle}} - \frac{ {\langle X \rangle}^2 }
		{ {\langle N \rangle}^2} }$
	\end{center}
	where:
	\begin{center}
		$X = \sum_{i=1}^N {p_T}_i \hspace{1cm} X_{2} = \sum_{i=1}^N \left ( {p_T}_i^2 \right )$
	\end{center}
	N is the number of charged particles in a given event and $\langle N \rangle$ is the mean multiplicity
        of the event sample.

	In order to obtain all these event quantities, two histograms are prepared:
	the one-dimensional distribution of $X_{2}$ and the two-dimensional distribution
        of $N$ vs. $X$.

	\section{Corrections}
\begin{figure}[!hbt]
		\centering
		\begin{subfigure}[b]{0.3\textwidth}
			\includegraphics[width=\textwidth]{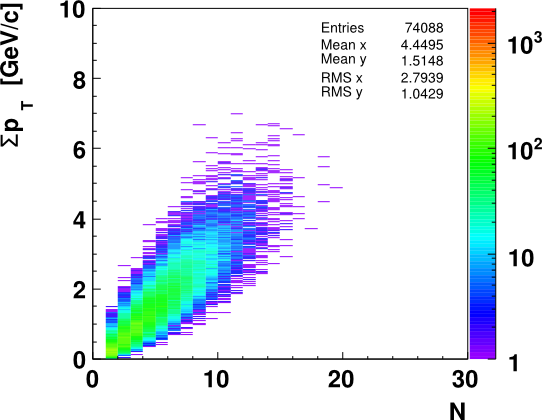}
		\end{subfigure}
		\hfill
		\begin{subfigure}[b]{0.3\textwidth}
			\includegraphics[width=\textwidth]{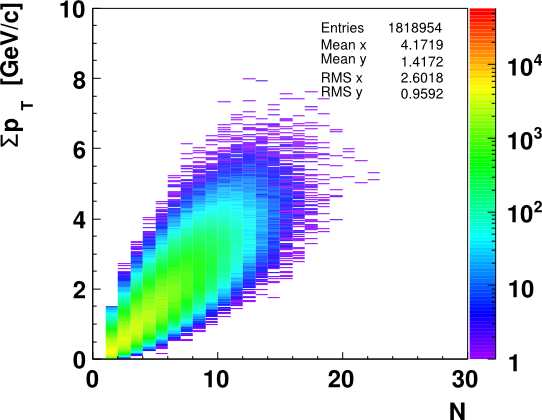}
		\end{subfigure}
		\hfill
		\begin{subfigure}[b]{0.3\textwidth}
			\includegraphics[width=\textwidth]{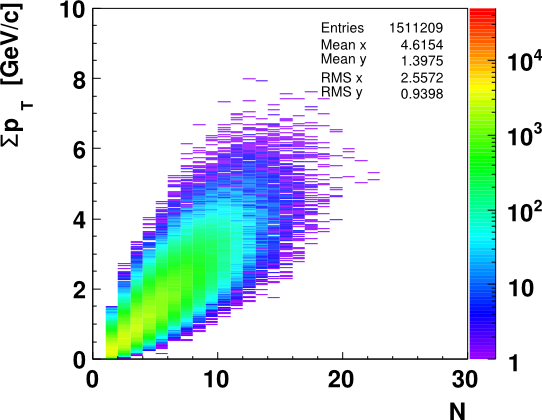}
		\end{subfigure}
		\caption{
			$N$ vs $X$ for 158 GeV/c target-removed (left), target-inserted (middle)
			and target-inserted	corrected for non-target interactions (right)
		}
		\label{fig:ETsubtract}
	\end{figure}

	Contamination by non-target interactions is subtracted as follows (Fig.~\ref{fig:ETsubtract}). NA61/SHINE records data with
	both target inserted and with no target. Then the correction is done by subtracting
	normalized histograms for data without the target (left) from the corresponding histograms for
	the target-inserted data (middle). The result is shown by the right plot.

\begin{figure}[!hbt]
		\centering
		\begin{subfigure}[b]{0.3\textwidth}
			\includegraphics[width=\textwidth]{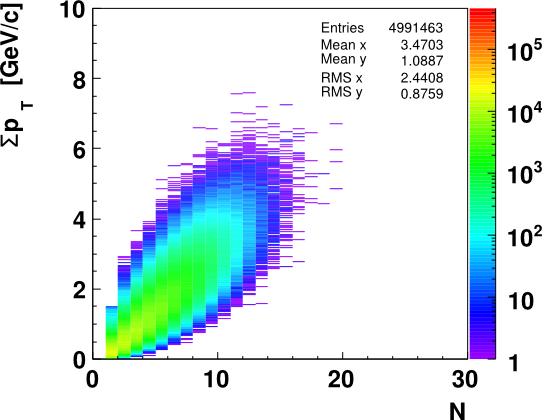}
		\end{subfigure}
		\hfill
		\begin{subfigure}[b]{0.3\textwidth}
			\includegraphics[width=\textwidth]{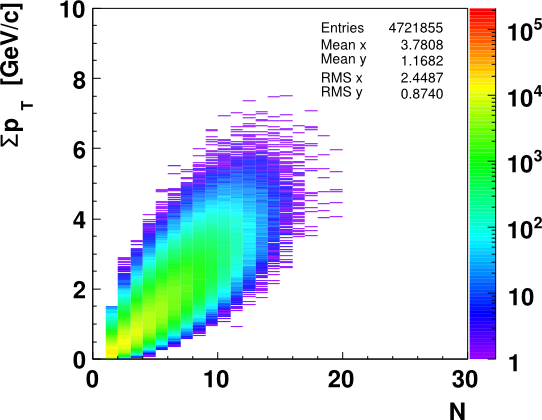}
		\end{subfigure}
		\hfill
		\begin{subfigure}[b]{0.3\textwidth}
			\includegraphics[width=\textwidth]{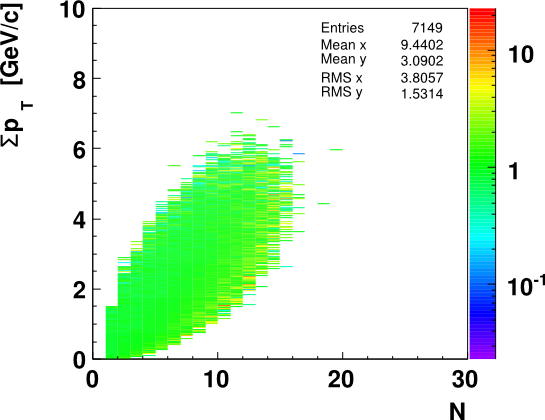}
		\end{subfigure}
		\caption{
			$N$ vs $X$ for 158 GeV/c VENUS reconstructed (left), VENUS generated (middle) and \\
			inverse correction factor (right).
		}
		\label{fig:MCgenrec}
	\end{figure}

	Eventually, $\Phi_{p_T}$ will be also corrected for detector effects. The inverse correction factor
	(Fig.~\ref{fig:MCgenrec} right)will be obtained by dividing corresponding histograms of the reconstructed
	VENUS data (middle) by pure generated VENUS (left).

	Then the data, after subtraction of the non-target interaction contamination, will be corrected
	by dividing the histograms by the inverse correction factor.

	\section{$\mathsf{\Phi_{p_T}}$ results}
\begin{figure}[!hbt]
		\centering
		\begin{subfigure}[b]{0.24\textwidth}
			\includegraphics[width=\textwidth]{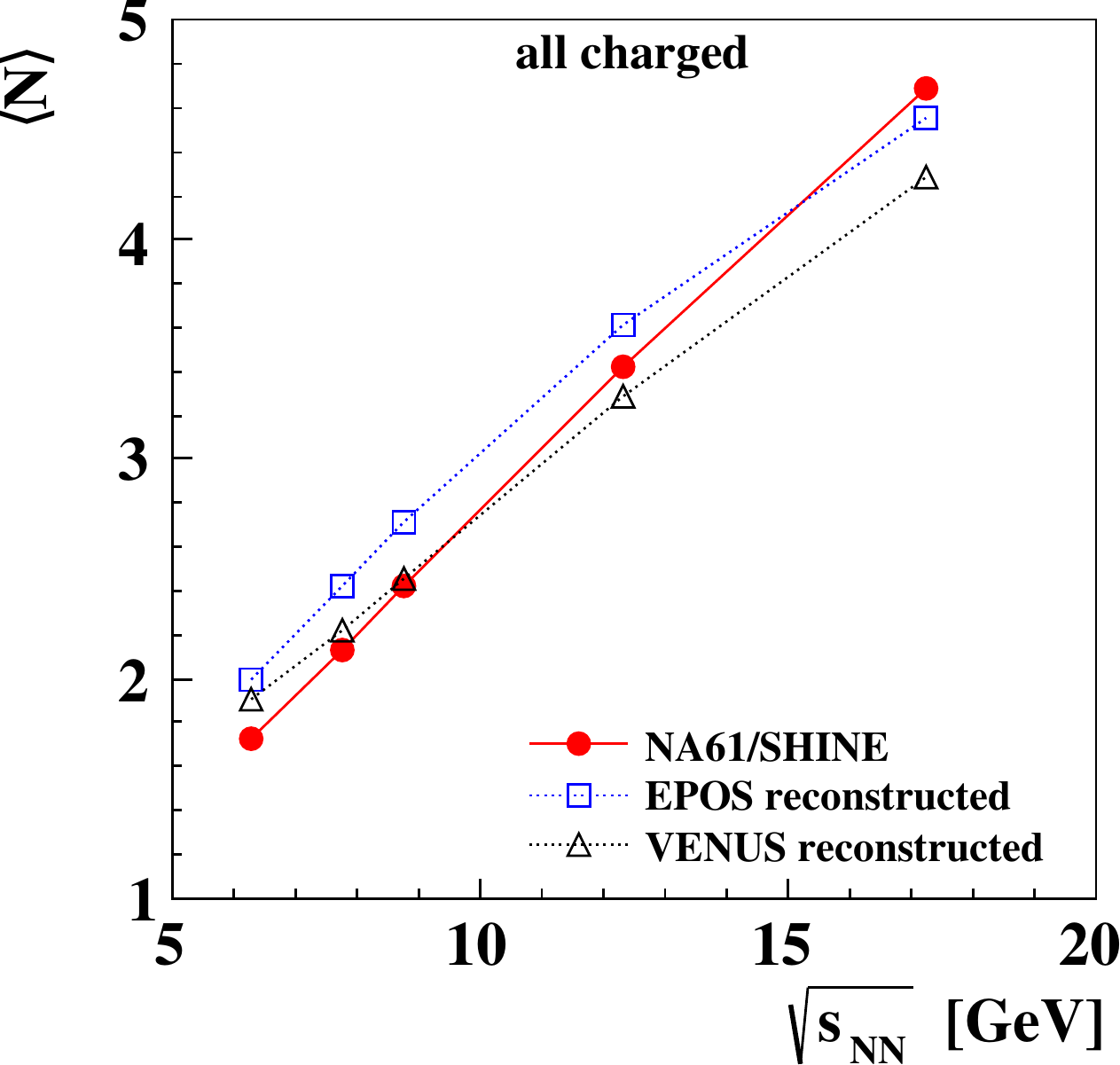}
		\end{subfigure}
		\hfill
		\begin{subfigure}[b]{0.24\textwidth}
			\includegraphics[width=\textwidth]{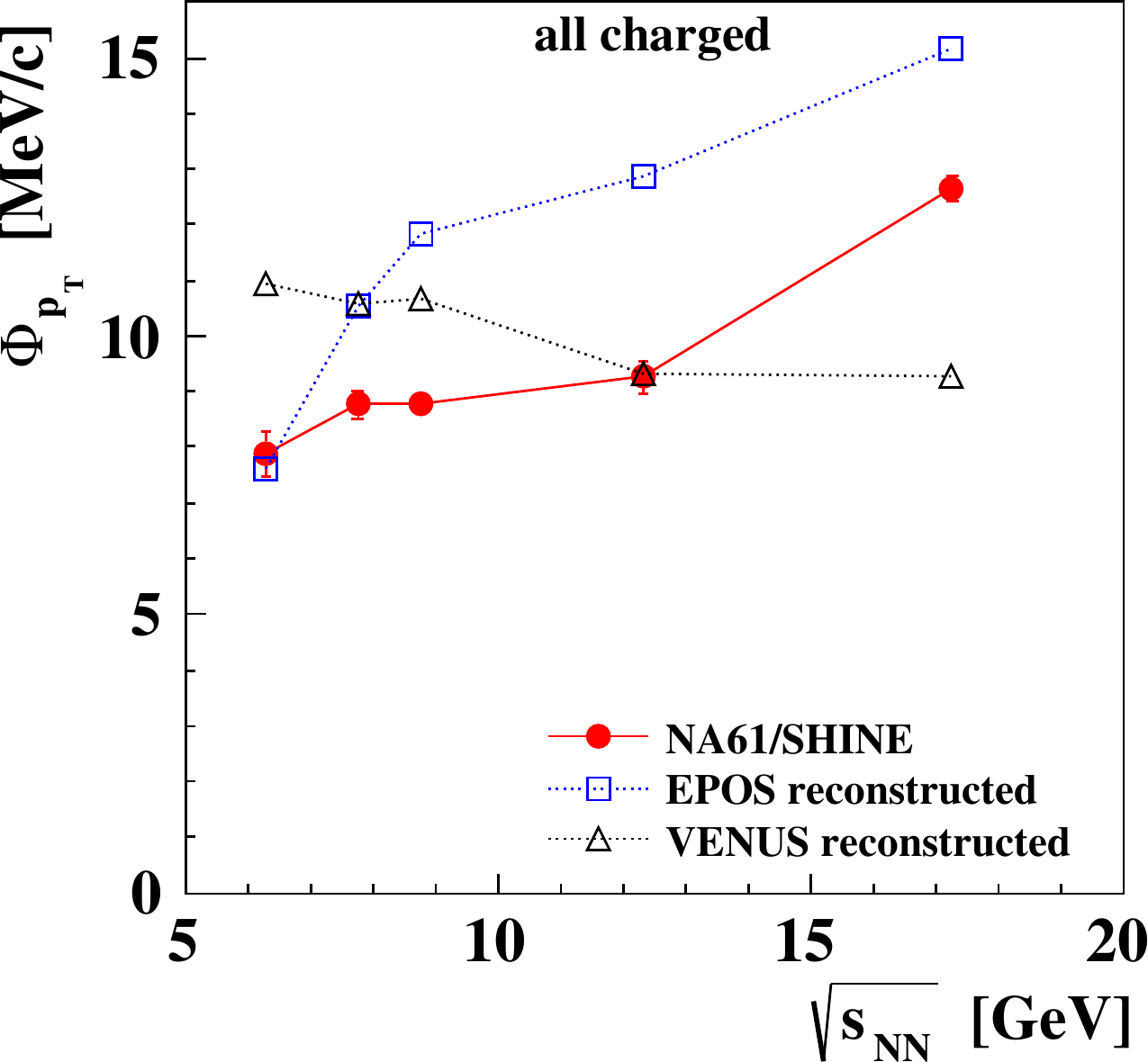}
		\end{subfigure}
		\hfill
		\begin{subfigure}[b]{0.24\textwidth}
			\includegraphics[width=\textwidth]{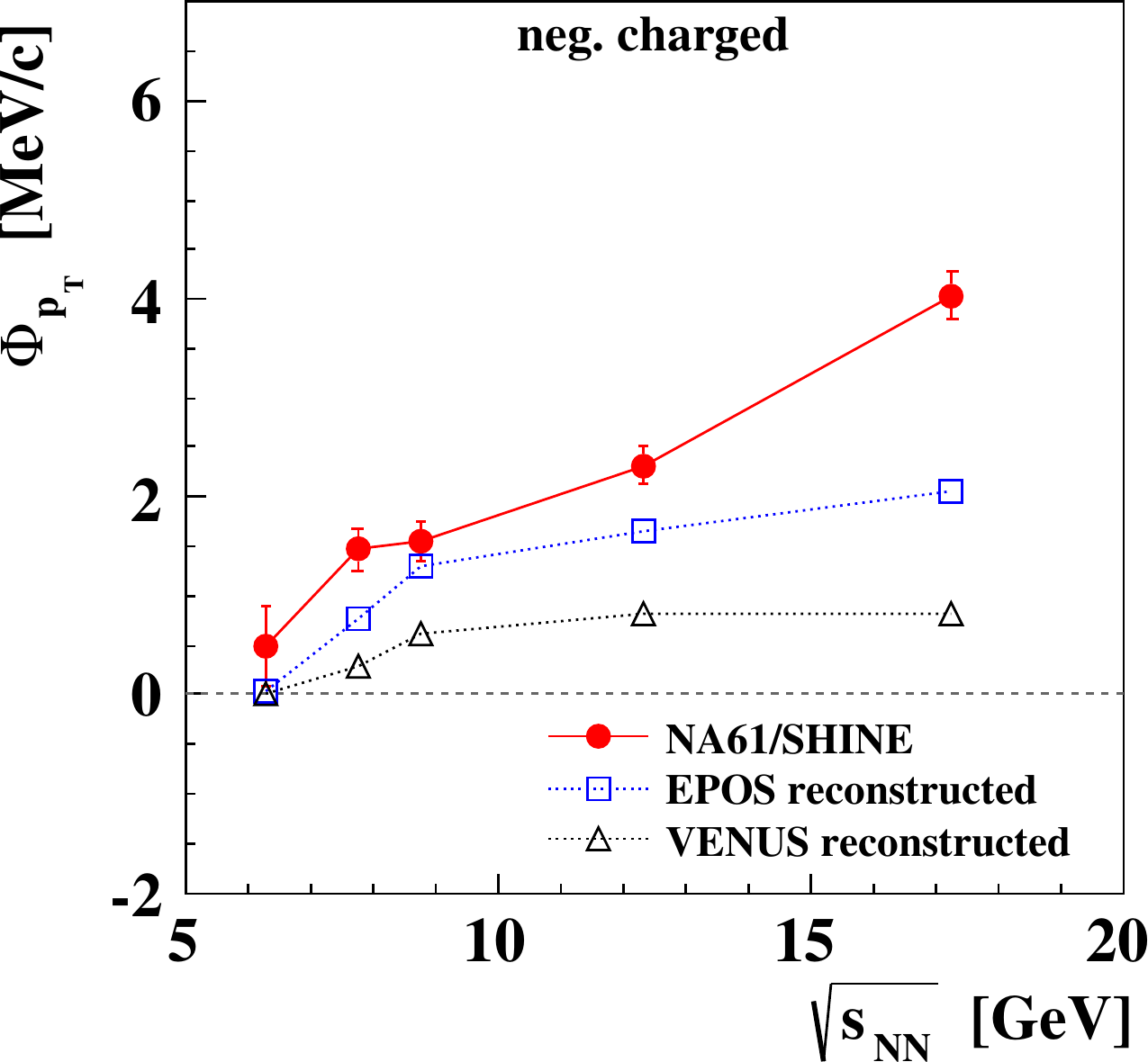}
		\end{subfigure}
		\hfill
		\begin{subfigure}[b]{0.24\textwidth}
			\includegraphics[width=\textwidth]{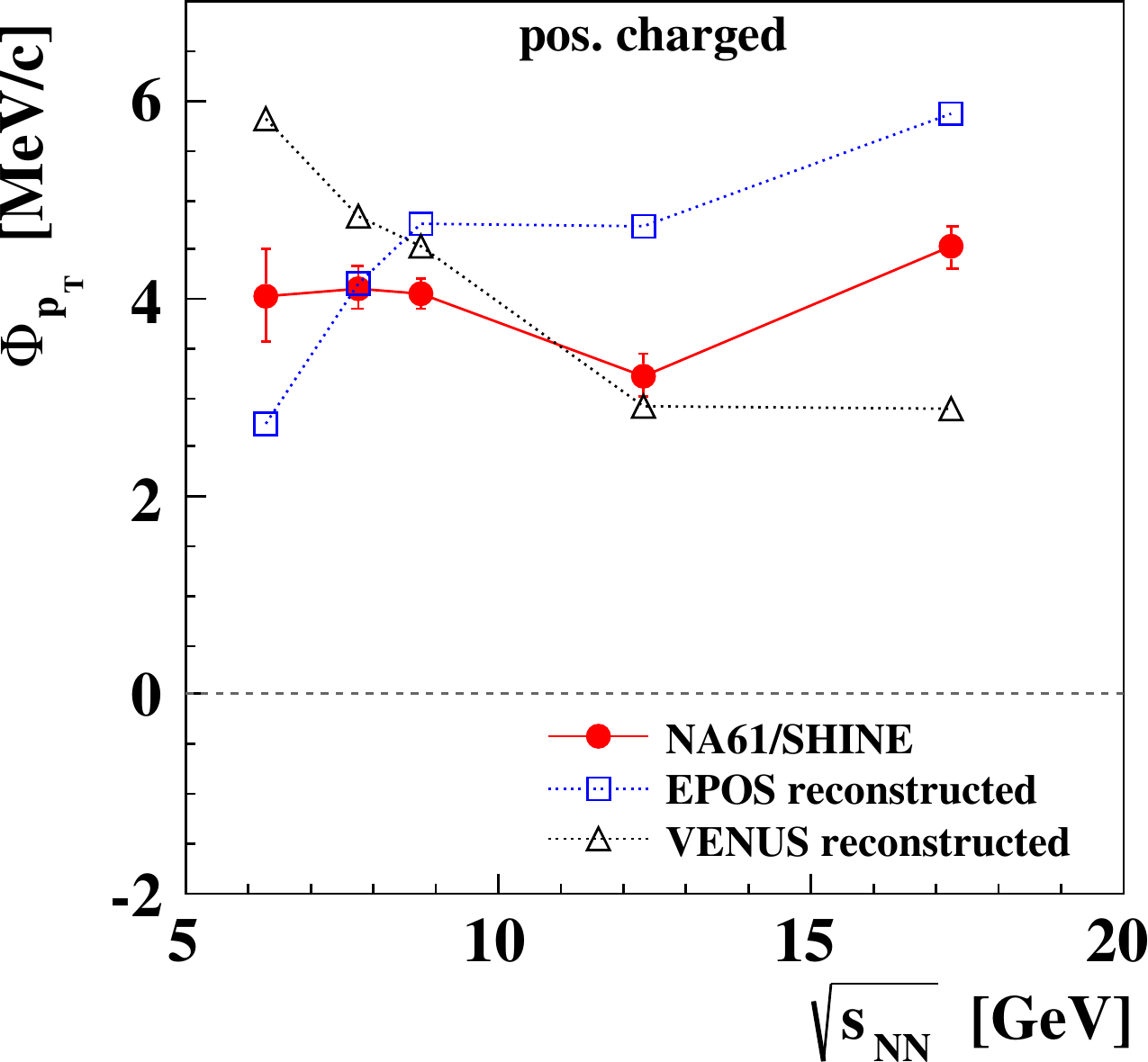}
		\end{subfigure}
		\caption{
			Mean multiplicity (left) and $\Phi_{p_T}$ for the p+p NA61/SHINE energy scan.
		}
		\label{fig:PhiResults}
	\end{figure}

\vspace{-0.2cm}
	Results, Fig.~\ref{fig:PhiResults}, are only corrected for non-target interactions. Study of corrections for detector
	effects and feed-down is still ongoing. Results from inelastic p+p interactions were collected
	at beam momenta: 20, 31, 40, 80, 158 GeV/c.

	In order to compare $\Phi_{p_T}$ from NA61/SHINE for p+p collisions and published NA49 data on central Pb+Pb interactions, additional NA49 acceptance cuts
	had to be applied:
	\begin{itemize}
		\item because of high density of tracks, analysis in NA49 was limited to the forward rapidity region ($1.1 < y < 2.6$),
		\item the smallest $\phi$ acceptance (20 GeV/c) was used for all energies.
	\end{itemize}
	Due to the additional cuts (mainly restricted rapidity) $\Phi_{p_T}$ for p+p collisions decreased.

\begin{figure}[!hbt]
		\centering
		\begin{subfigure}[b]{0.3\textwidth}
			\includegraphics[width=\textwidth]{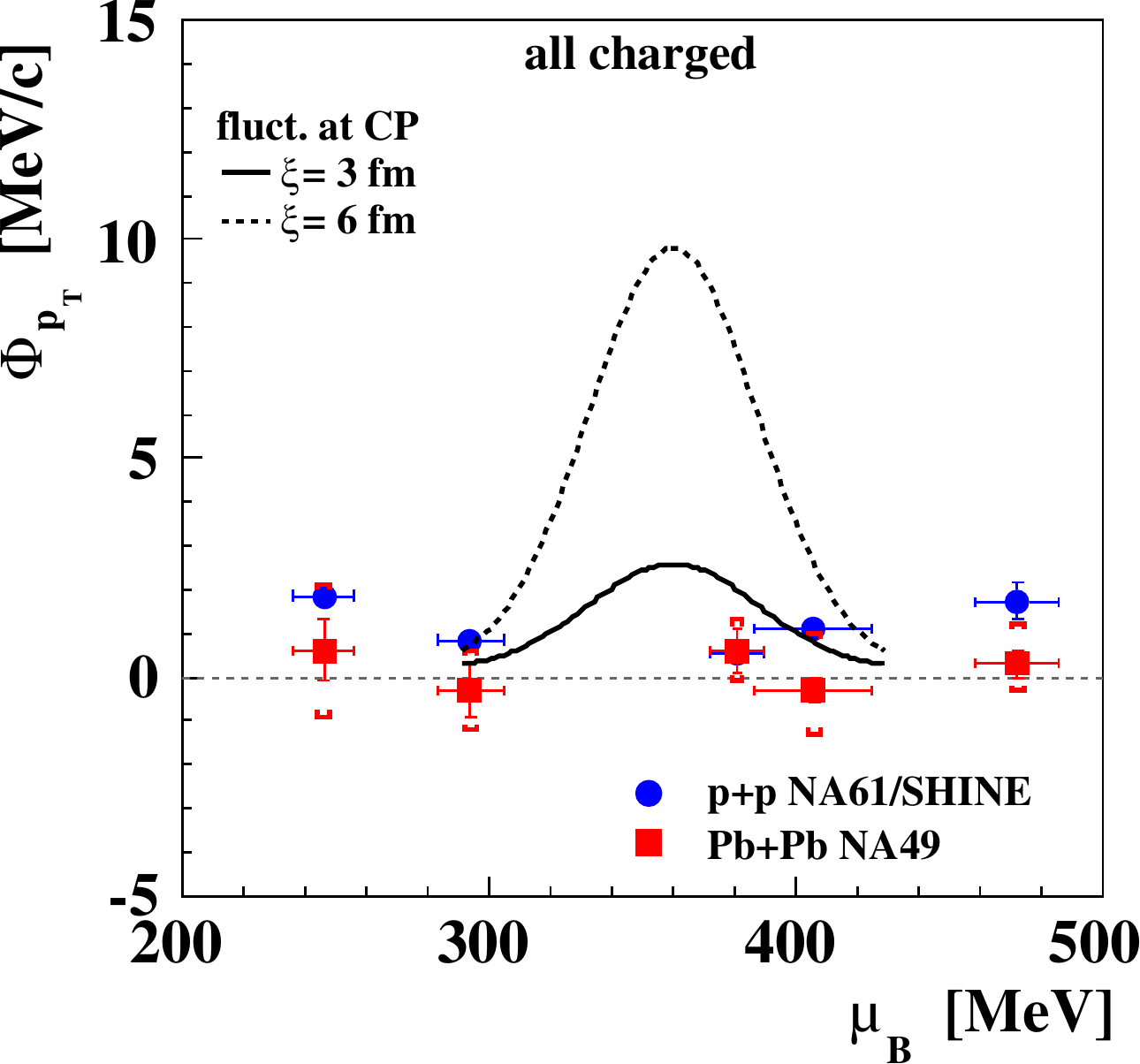}
		\end{subfigure}
		\quad
		\begin{subfigure}[b]{0.3\textwidth}
			\includegraphics[width=\textwidth]{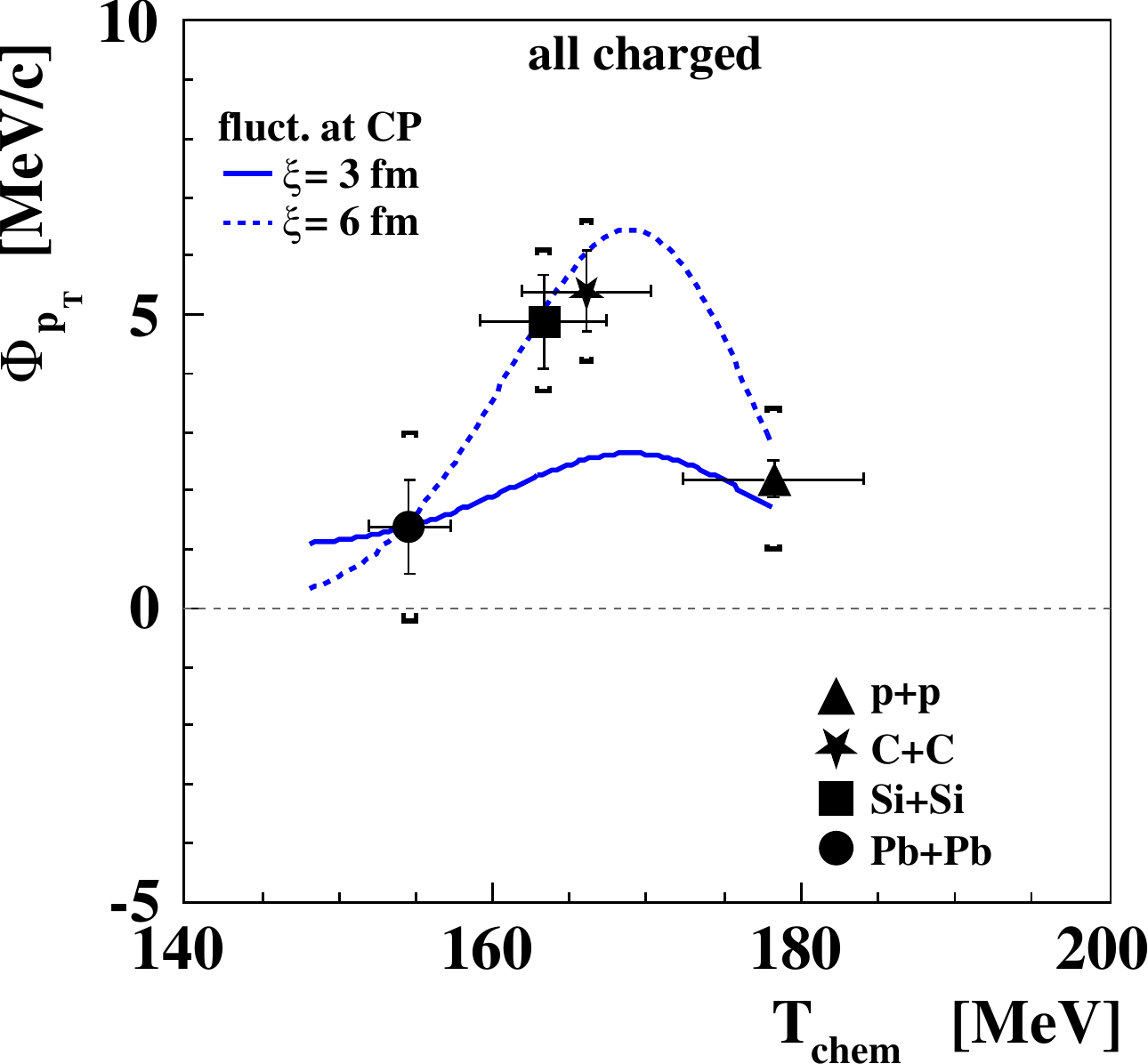}
		\end{subfigure}
		\caption{
			Comparison of NA61 p+p and NA49 central Pb+Pb results in common acceptance with
			predictions for the critical point (left). System-size scan published by NA49 with
			predictions for the critical point (right).
		}
		\label{fig:PhiComparison}
	\end{figure}

	The collision energy dependence of $\Phi_{p_T}$ for p+p and central Pb+Pb collisions (Fig.~\ref{fig:PhiComparison} left) does not show any anomalies
	which might be expected for the critical point. A hint of a signal is, however, observed for medium size systems (right).
	NA61/SHINE will record Ar+Ca and Xe+La data in 2014/15 which will allow a systematic study.

	\section{Two-particle correlations}
\begin{figure}[!hbt]
		\centering
		\begin{subfigure}[b]{0.49\textwidth}
			\includegraphics[width=\textwidth]{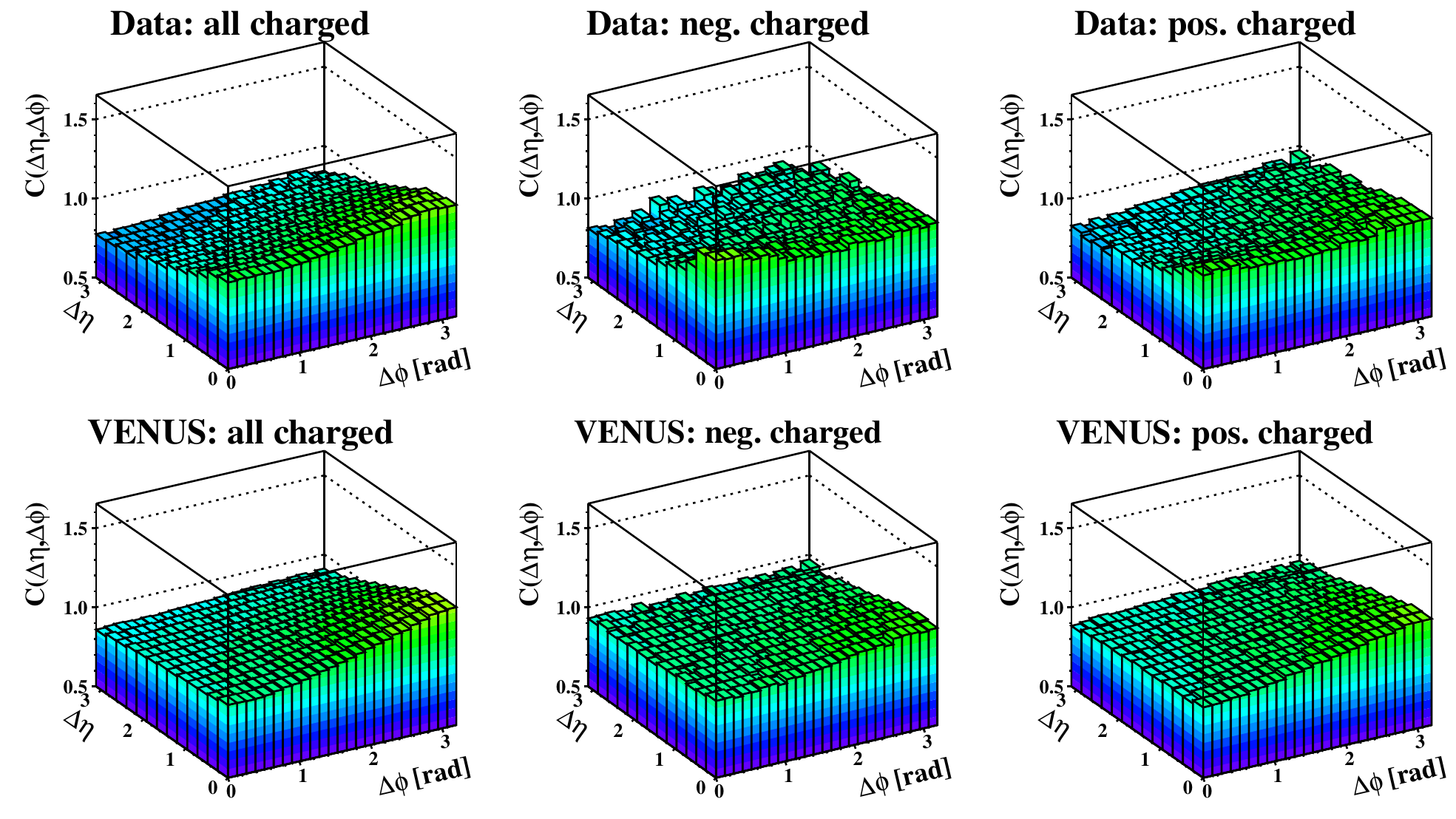}
		\end{subfigure}
		\hfill
		\begin{subfigure}[b]{0.49\textwidth}
			\includegraphics[width=\textwidth]{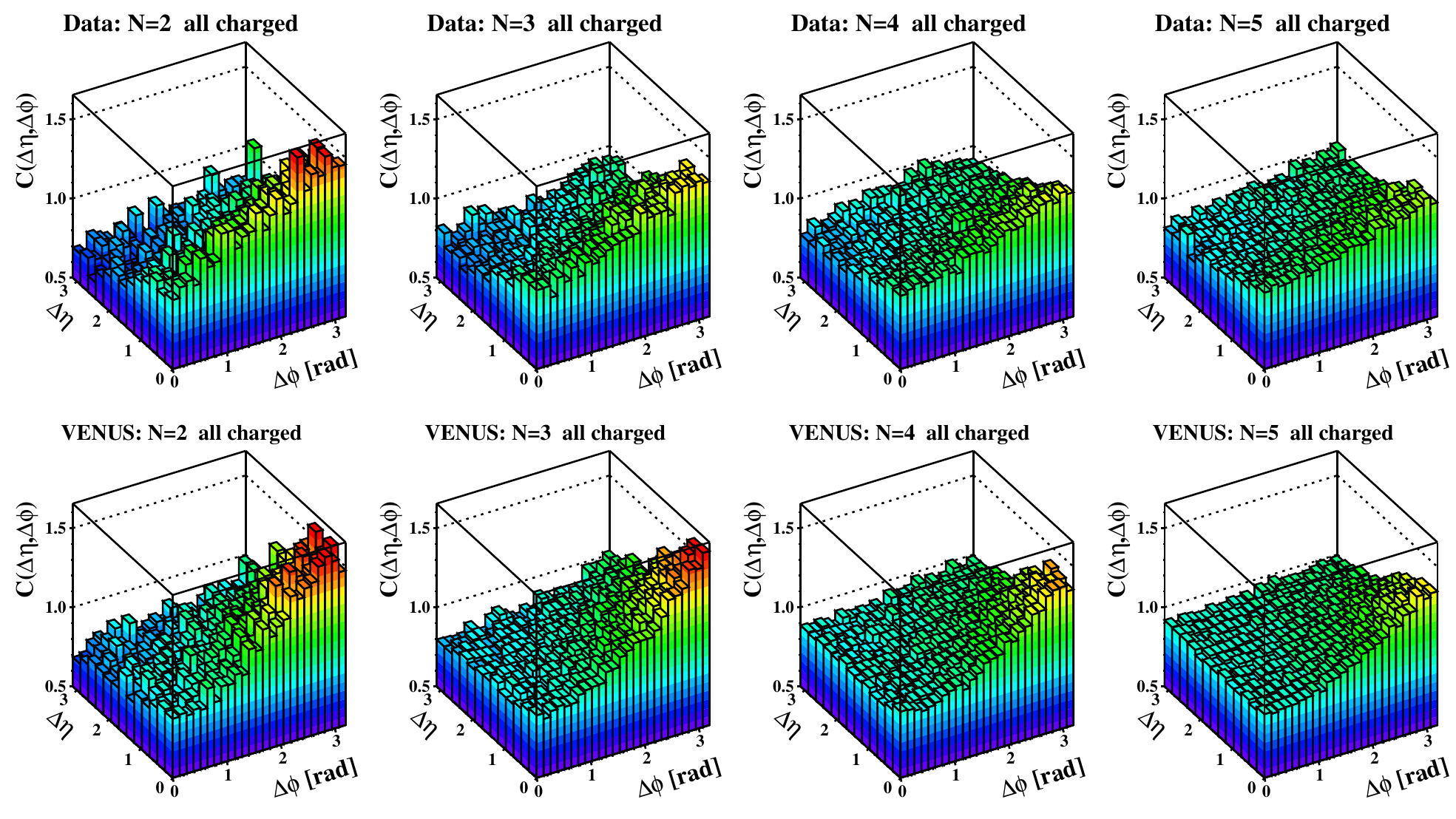}
		\end{subfigure}
		\caption{
			$\Delta\eta\Delta\phi$ correlations for all events (left) and for multiplicities of 2, 3, 4 and 5 (right)
			for all charged particles
		}
		\label{fig:CorrAll}
	\end{figure}

	Two-particle correlations were studied for inelastic p+p interactions at 158 GeV/c
	beam momentum as a function of the difference in pseudo-rapidity $\eta$
	and azimuthal angle $\phi$ between two particles in the same event:
	\begin{center}
		$\Delta\eta = |{\eta}_1 - {\eta}_2| \hspace{1cm} \Delta\phi = |{\phi}_1 - {\phi}_2|$
	\end{center}
	The correlation function is defined as:
	\begin{center}
		$C(\Delta\eta,\Delta\phi)=\frac{N_{mixed}^{pairs}}{N_{data}^{pairs}}\frac{S(\Delta\eta,\Delta\phi)}
		{M(\Delta\eta,\Delta\phi)}$
		\\
		$S(\Delta\eta,\Delta\phi)=\frac{d^2N^{signal}}{d \Delta \eta d \Delta \phi}$;
		\hspace{0.1cm}
		$M(\Delta\eta,\Delta\phi)=\frac{d^2N^{mixed}}{d \Delta \eta d \Delta \phi}$
	\end{center}
	The correlation function was folded around $\Delta\phi = 0$, i.e. if $\Delta\phi > \pi$ then $\Delta\phi = 2\pi - \Delta\phi$.
	$\eta$ was transformed from LAB to CMS assuming the $\pi$ mass. Correlations for all events
	are shown in Fig.~\ref{fig:CorrAll} (left).

	Correlations were also calculated for events with multiplicities of 2, 3, 4 and 5 for all charged (Fig.~\ref{fig:CorrAll} right),
	positively charged (below, left) and negatively charged particles (below, right).

	\begin{figure}[!hbt]
		\centering
		\begin{subfigure}[b]{0.49\textwidth}
			\includegraphics[width=\textwidth]{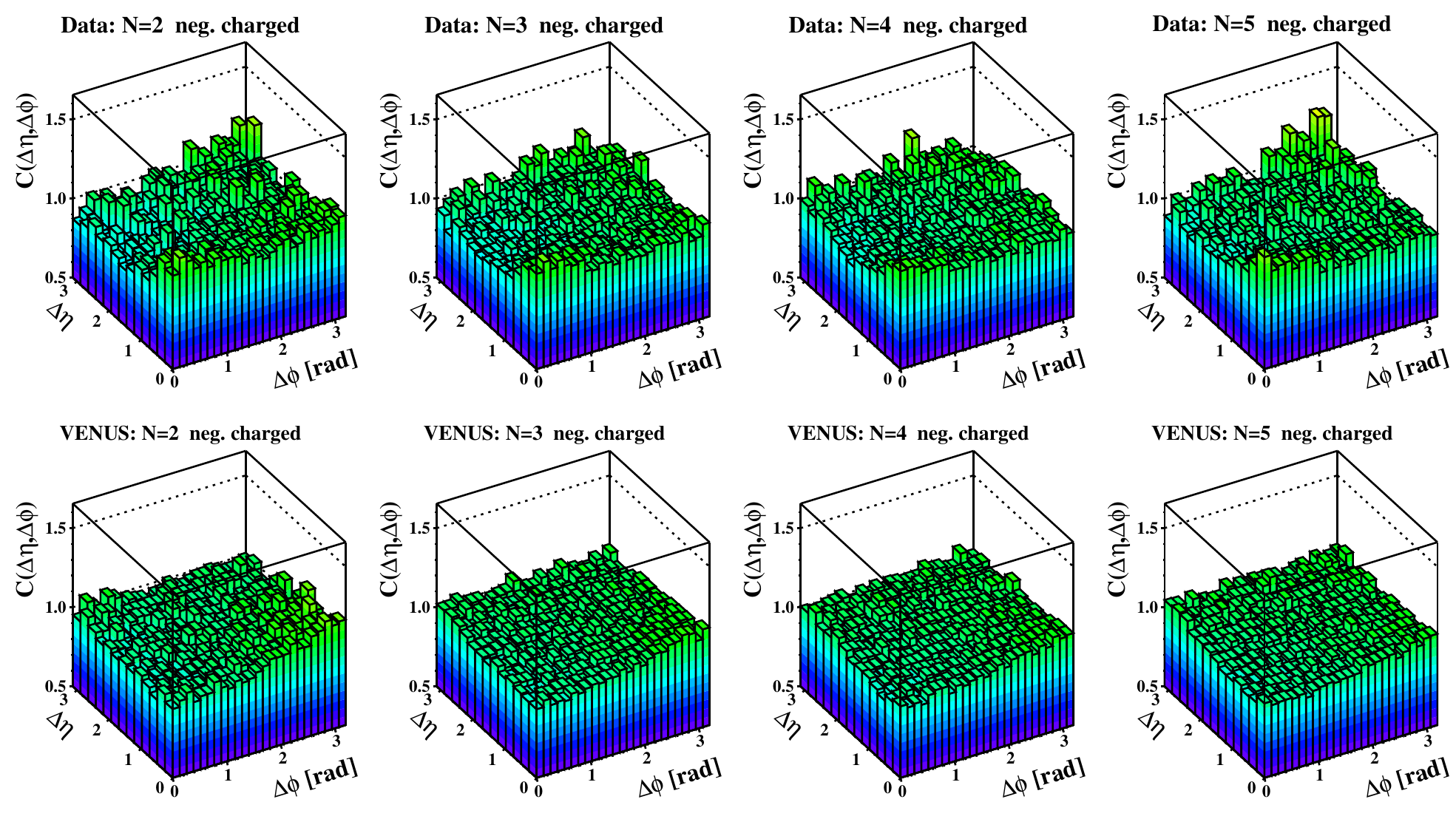}
		\end{subfigure}
		\hfill
		\begin{subfigure}[b]{0.49\textwidth}
			\includegraphics[width=\textwidth]{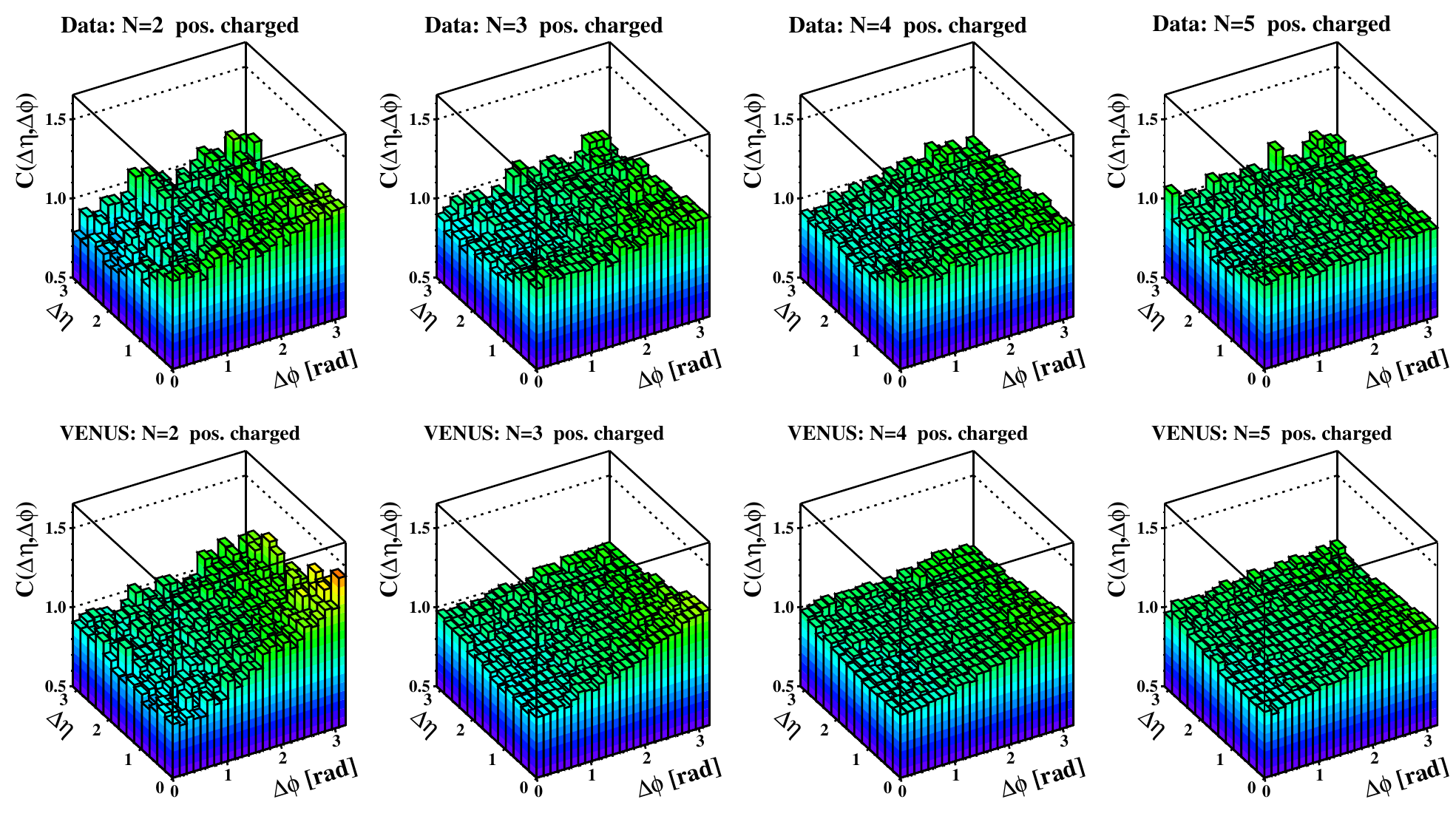}
		\end{subfigure}
		\caption {
			$\Delta\eta\Delta\phi$ correlations for events with multiplicities 2, 3, 4 and 5 for negatively (left)
			and positively (right) charged particles
		}
		\label{fig:CorrNegPos}
	\end{figure}

	Results from NA61 data (upper distributions) show several structures:
	\begin{itemize}
		\item A significant maximum for all and positively charged particles at $(\Delta\eta,\Delta\phi)=(0,\pi)$
		  probably coming from resonance decays and momentum conservation. This maximum is better visible
		  for selected  multiplicities (see Fig.~\ref{fig:CorrNegPos}).
		\item A lower but also visible enhancement at $(0,0)$ coming from Coulomb interactions and Bose-Einstein correlation (quantum) effects.
	\end{itemize}
	Results from the VENUS model (bottom distributions) show similar structures except the hill at $(0,0)$,
	because VENUS does not simulate Coulomb and quantum effects.

	Two-particle correlations in $\Delta\eta\Delta\phi$ were also studied at RHIC \cite{STAR} and LHC \cite{CMS}.

	\section{Summary}
	A new method of correcting results of fluctuations for experimental and physics biases was introduced.
	$\Phi_{p_T}$ results corrected for non-target interactions for inelastic p+p interactions at beam momenta:
	20, 31, 40, 80 and 158 GeV/c were presented. Fully corrected results will be shown soon.

	Two-particle correlations were obtained for inelastic p+p interactions at~158~GeV/c both for NA61 data and
	the VENUS model. The analysis showed several structures. They depend on the multiplicity selection and may
	be related to Coulomb interactions, quantum effects, resonance decays and conservation laws.

	\section*{Acknowledgements}
	This work was supported by the polish National Science Centre grant DEC-2011/03/B/ST2/02617 and grant
	2012/04/M/ST2/00816.

\end{document}